\documentclass[12pt,thmsa]{article}
\usepackage{amsfonts}
\usepackage{amssymb}
\usepackage{graphicx}

\input{tcilatex}
\begin{document}

\begin{center}
{\large Entangled states generation of two atoms interacting with a cavity
field}

{\small F. N. M. Al-Showaikh}$^{1}${\small , N. Metwally}$^{1,2}${\small \
and M. Abdel-Aty}$^{1,3}$

$^{1}${\small Mathematics Department, College of Science, Bahrain
University, Bahrain}

$^{2}${\small Mathematics Department, Faculty of Science, South Valley
University, Aswan, Egypt}

$^{3}${\small Mathematics Department, Faculty of Science, Sohag University,
82524 Sohag, Egypt}

\smallskip
\end{center}

We classify different classes of entangled states arise in a
two-qubit system. Some of these classes are of Bell's state types,
while others are of the Werner's state types. The degree of
entanglement is quantified for different values of the atomic and
the cavity parameters. We show that it is possible to generate
entangled state with high degree of entanglement by controlling
the detuning and the number of photon inside the cavity.

\section{Introduction}

Although physicists have been able to entangle up to eight qubits
at the same time -- whereas recently interest has been focused on
just entangling two qubits \cite{ste06}. They insist
superconducting qubits are a viable approach towards quantum
computing. Before this work, a number of applications have been
studied experimentally and theoretically, see for example
\cite{Ph}. The coupling of multi two-level atoms to a single
quantized radiation mode has been solved. Finite algorithms for
the eigenfrequencies were given for the case of resonance
\cite{10}. The coupling of a single two-level atom to a number
equal-frequency modes has been used as a simple model of decay
processes \cite{11}. However, as a result of the development of
low-temperature high-$Q$ cavities, and the use of Rydberg atoms,
the idealized situation of a single two-level atom interacting
with a single-mode,quantized radiation field in a lossless cavity
has been experimentally realized \cite{12}. These advances in turn
have certainly provided incentive for extending and generalizing
this model. Thus many authors have been encouraged and stimulated
to modify the original model and to generalize it into different
directions.

Nowadays, new applications of the atomic systems have been appeared e.g. in
quantum information \cite{You,Su,Metwally2,Metwally} and computations \cite%
{Bige}. The most important phenomena in the atomic systems which is needed
in quantum information tasks is the entanglement. So it is of a great
importance to investigate how one can generate entanglement between atoms.

In this contribution, we consider an atomic system interacts with a cavity
mode field in a Fock state. This atomic system is prepared in excited or
ground state. Due to this interaction, there are different classes of
entangled states between the two atoms have been generated. Then we quantify
the degree of entanglement contained in these entangled states. This article
is organized as follows: In Sec.2, we describe our model and give its
analytical solution. Investigating the dynamics of the population is shown
in Sec.3. Then we quantify the degree of entanglement for different initial
setting and different values of the field and atomics parameter as shown in
Sec.4. Finally, we give our conclusion in Sec.5.

\section{The Model and its Solution}

The Hamiltonian which describes a system of two two-level atoms, each
consisting of states $\bigl|e\bigr\rangle$ and $\bigl|g\bigr\rangle$ coupled
to a single mode radiation field, in the rotating wave approximation is
given by
\begin{equation}
H=\omega (a^{\dagger }a+\sum_{i=1}^{2}\sigma
_{z}^{i})+\sum_{i=1}^{2}[\lambda _{i}(a^{\dagger }\sigma _{-}^{i}+\sigma
_{+}^{i}a)]
\end{equation}%
where $a(a^{\dagger })$ is the annihilation (creation) operator of the field
mode, $\sigma _{\pm }^{i}$ and $\sigma _{z}^{i}$ are the atomic raising,
lowering and inversion operators of the two atoms. The parameter $\lambda
_{i}$ is the atom-field coupling constant and $\omega $, is the atomic
transitions and the field mode frequency. The first term in Eq.$(1)$
represents the free-field and the non-interacting atoms, while the second
term stands for the interaction Hamiltonian $H_{int}$. This model has been
solved analytically for some special cases \cite{Dung} and for a general
case \cite{Obada}. For simplicity, we consider the case of identical atoms
i.e. $\lambda _{1}=\lambda _{2}$. Assume that the cavity field is initially
prepared in a coherent state, $\bigl|\psi (0)\bigr\rangle_{f}=\bigl|n%
\bigr\rangle$ and the two atoms are in a superposition states of their
ground and exited states $\bigl|\psi (0)\bigr\rangle_{1,2}=a_{i}\bigl|g_{i}%
\bigr\rangle+b_{i}\bigl|e_{i}\bigr\rangle,i=1,2$, where $1$ stands for the
first atom and $2$ for the second atom. The initial state of the system is
assumed to be,
\begin{equation}
\bigl|\psi _{0}\bigr\rangle=\bigl(c_{0}^{(1)}\bigl|g_{1}g_{2}\bigr\rangle%
+c_{0}^{(2)}\bigl|e_{1}g_{2}\bigr\rangle+c_{0}^{(3)}\bigl|g_{1}e_{2}%
\bigr\rangle+c_{0}^{(4)}\bigl|e_{1}e_{2}\bigr\rangle\bigr)\otimes \bigl|n%
\bigr\rangle
\end{equation}%
where
\begin{equation}
c_{0}^{(1)}=a_{1}a_{2},\quad c_{0}^{(2)}=a_{2}b_{1},\quad
c_{0}^{(3)}=a_{1}b_{2},\quad c_{0}^{(4)}=b_{1}b_{2}  \label{inCof}
\end{equation}%
with $\mid a_{i}\mid ^{2}+\mid b_{i}\mid ^{2}=1$ for $i=1,2$. In the
invariant sub-space of the global system, we can consider a set of complete
basis of the qubit-field system as $\bigl|ee,n\bigr\rangle,\bigl|eg,n+1%
\bigr\rangle,\bigl|ge,n+1\bigr\rangle$ and $\bigl|gg,n+2\bigr\rangle$. The
time evolution of the density operator of the system is given by
\begin{equation}
\varrho _{cf}(t)=\mathcal{U}(t)\{\varrho _{a}(0)\otimes \varrho _{f}(0)\}%
\mathcal{U}^{\dagger }(t),  \label{Denst}
\end{equation}%
where $\mathcal{U}(t)=\exp \left( -i\hat{H}t/\bar{h}\right) $ is the unitary
operator, its components are given by,
\begin{eqnarray}
\mathcal{U}_{11}(t) &=&\sum_{i=1}^{3}{(-1)^{i+1}\alpha _{i}e^{-i\mu
_{1}t}[\mu _{i}(\Delta +\mu _{i})-2\beta ^{2}]},  \nonumber  \label{Unit} \\
\mathcal{U}_{12}(t) &=&\gamma \sum_{i=1}^{3}{(-1)^{i+1}e^{-i\mu
_{i}t}(\Delta +\mu _{i})},  \nonumber \\
\mathcal{U}_{13}(t) &=&\mathcal{U}_{12}(t),\quad \mathcal{U}_{14}(t)=2\beta
\gamma \sum_{i=1}^{3}{(-1)^{i+1}\alpha _{i}e^{-i\mu _{i}t}},  \nonumber \\
\mathcal{U}_{21}(t)&=&\mathcal{U}_{12}(t)
\nonumber\\
\mathcal{U}_{22}(t) &=&\sum_{i=1}^{3}{(-1)^{i+1}\frac{\alpha _{i}}{\mu _{i}}%
e^{-i\mu _{i}t}\Bigl[(\beta ^{2}(\Delta -\mu _{i})-(\Delta +\mu
_{i}))(\gamma ^{2}+\mu _{i}(\Delta -\mu _{i}))\Bigr]}-\frac{\Delta (\beta
^{2}-\gamma ^{2})}{\mu _{1}\mu _{2}\mu _{3}},  \nonumber \\
\mathcal{U}_{23}(t) &=&-\sum_{i=1}^{3}{(-1)^{i+1}\frac{\alpha _{i}}{\mu _{i}}%
e^{-i\mu _{i}t}\Bigl[\beta ^{2}(\Delta -\mu _{i})-\gamma ^{2}(\Delta +\mu
_{i})\Bigr]}+\frac{\Delta (\beta ^{2}-\gamma ^{2})}{\mu _{1}\mu _{2}\mu _{3}}%
,  \nonumber \\
\mathcal{U}_{24}(t) &=&-\beta \sum_{i=1}^{3}{(-1)^{i+1}\alpha _{i}e^{-i\mu
_{i}t}(\Delta -\mu _{i})},  \nonumber \\
\mathcal{U}_{34}(t) &=&-\beta \sum_{i=1}^{3}{(-1)^{i+1}\alpha _{i}e^{-i\mu
_{i}t}(\Delta -\mu _{i})},  \nonumber \\
\mathcal{U}_{31}(t) &=&\mathcal{U}_{13}(t),\quad \mathcal{U}_{32}(t)=%
\mathcal{U}_{23}(t),\quad \mathcal{U}_{33}(t)=\mathcal{U}_{22}(t),  \nonumber
\\
\mathcal{U}_{41}(t) &=&\mathcal{U}_{14}(t),\quad \mathcal{U}_{42}(t)=%
\mathcal{U}_{24}(t),\quad \mathcal{U}_{43}(t)=\mathcal{U}_{34}(t),  \nonumber
\\
\mathcal{U}_{44}(t) &=&-\sum_{i=1}^{3}{(-1)^{i+1}\alpha _{i}e^{-i\mu _{i}t}%
\Bigl[2\gamma ^{2}+\mu _{i}(\Delta -\mu _{i})\Bigr]},
\end{eqnarray}%
where $\gamma =\sqrt{n+1},\beta =\sqrt{n+2}$, $\alpha _{1}=(\mu _{12}\mu
_{13})^{-1},~\alpha _{2}=(\mu _{12}\mu _{23})^{-1},\alpha _{3}=(\mu _{13}\mu
_{23})^{-1}$, $\mu _{kj}=\mu _{k}-\mu _{j}$ and $\mu _{i}=\frac{2}{3}\kappa
\cos \theta _{i}$ with $\kappa =\sqrt{3(\Delta ^{2}+2(\beta ^{2}+\gamma
^{2}))}$ and $\theta _{1}=\frac{1}{3}\cos ^{-1}\left( -\frac{27\Delta }{%
\kappa ^{3}}\right) $,~ $\theta _{2}=\frac{2\pi }{3}+\theta _{1},~\theta
_{3}=\frac{2\pi }{3}+\theta _{2}$.

Since we are interested in discussing some properties of the charge qubits
system, we calculate the density matrix of the charged qubit by tracing out
the field i.e $\varrho_{ab}=tr_{f}\{\varrho_{cf}\}$ .

\begin{equation}  \label{DAtoms}
\rho_{atoms}= \left(
\begin{array}{cccc}
|A_n^{(1)}|^2 & A_n^{(1)}A_{n+1}^{*(2)} & A_n^{(1)}A_{n+1}^{*(3)} &
A_n^{(1)}A_{n+2}^{*(4)} \\
A_{n+1}^{(2)}A_{n}^{*(1)} & |A_n^{(2)}|^2 & A_{n}^{(2)}A_{n}^{*(3)} &
A_n^{(2)}A_{n+1}^{*(4)} \\
A_{n+1}^{(3)}A_{n}^{*(1)} & A_n^{(3)}A_{n}^{*(2)} & |A_{n}^{(3)}|^2 &
A_n^{(3)}A_{n+1}^{*(4)} \\
A_{n+2}^{(4)}A_{n}^{*(1)} & A_{n+1}^{(4)}A_{n}^{*(2)} &
A_{n+1}^{(4)}A_{n}^{*(3)} & |A_n^{(4)}|^2%
\end{array}
\right)
\end{equation}

with,
\begin{eqnarray}
A_{n}^{(1)} &=&\sum_{j=1}^{4}{\mathcal{U}_{1j}(n)A^{(j)}(0)},\quad
A_{n}^{(2)}=\sum_{j=1}^{4}{\mathcal{U}_{2j}(n)A^{(j)}(0)},\quad  \nonumber \\
A_{n}^{(3)} &=&\sum_{j=1}^{4}{\mathcal{U}_{3j}(n)A^{(j)}(0)},\quad
A_{n}^{(4)}=\sum_{j=1}^{4}{\mathcal{U}_{4j}(n)A^{(j)}(0)},
\end{eqnarray}%
where, $%
A^{(1)}(0)=b_{1}b_{2},~A^{(2)}(0)=b_{1}a_{2},~A^{(3)}(0)=a_{1}b_{2},~A^{(4)}(0)=a_{1}a_{2}.
$

\section{Entangled state}

Due to the interaction, the dynamical system behaves as an entangled state
for some time and as a product in another interval of time. In this section,
we show the effect of the field and the atomic parameters on the possibility
of generating entangled state with high degree of entanglement. For this
purpose, we plot the populations. In Fig.(1), we plot the populations for
atomic system is prepared initially in excited state. In this figure, we
investigate the dynamics of this phenomena for different values of the
detuning parameter. In Fig.(1a), we assume that $\Delta =0.1$ and $\bar{n}=0$%
, it is clear that the there are some entangled state are generated at
different time. As an example for $\tau >0$, the first partially entangled
state has the form
\begin{equation}
\bigl|\psi _{1}\bigr\rangle=\mu (\bigl|eg\bigr\rangle+\bigl|ge\bigr\rangle),
\end{equation}%
where $0<\mu \leq 2.9$. This class of state is similar to the class of the
Bell state \cite{Bell, Niel}, $\bigl|\psi ^{+}\bigr\rangle=\frac{1}{\sqrt{2}}(\bigl|01%
\bigr\rangle+\bigl|01\bigr\rangle)$. This class of entangled state
appears
along the time expect at some time where it turns to a separable states at $%
\tau \simeq 1.5,2.5,3.75,...$. Then as the time increases more, there is a
different class of partially entangled state is generated. This class could
be written as,
\begin{equation}
\bigl|\psi _{2}\bigr\rangle=\mu _{1}(\bigl|ee\bigr\rangle+\bigl|eg%
\bigr\rangle+\bigl|ge\bigr\rangle)
\end{equation}%
\begin{figure}[tbp]
\begin{center}
\includegraphics[width=18pc,height=12pc]{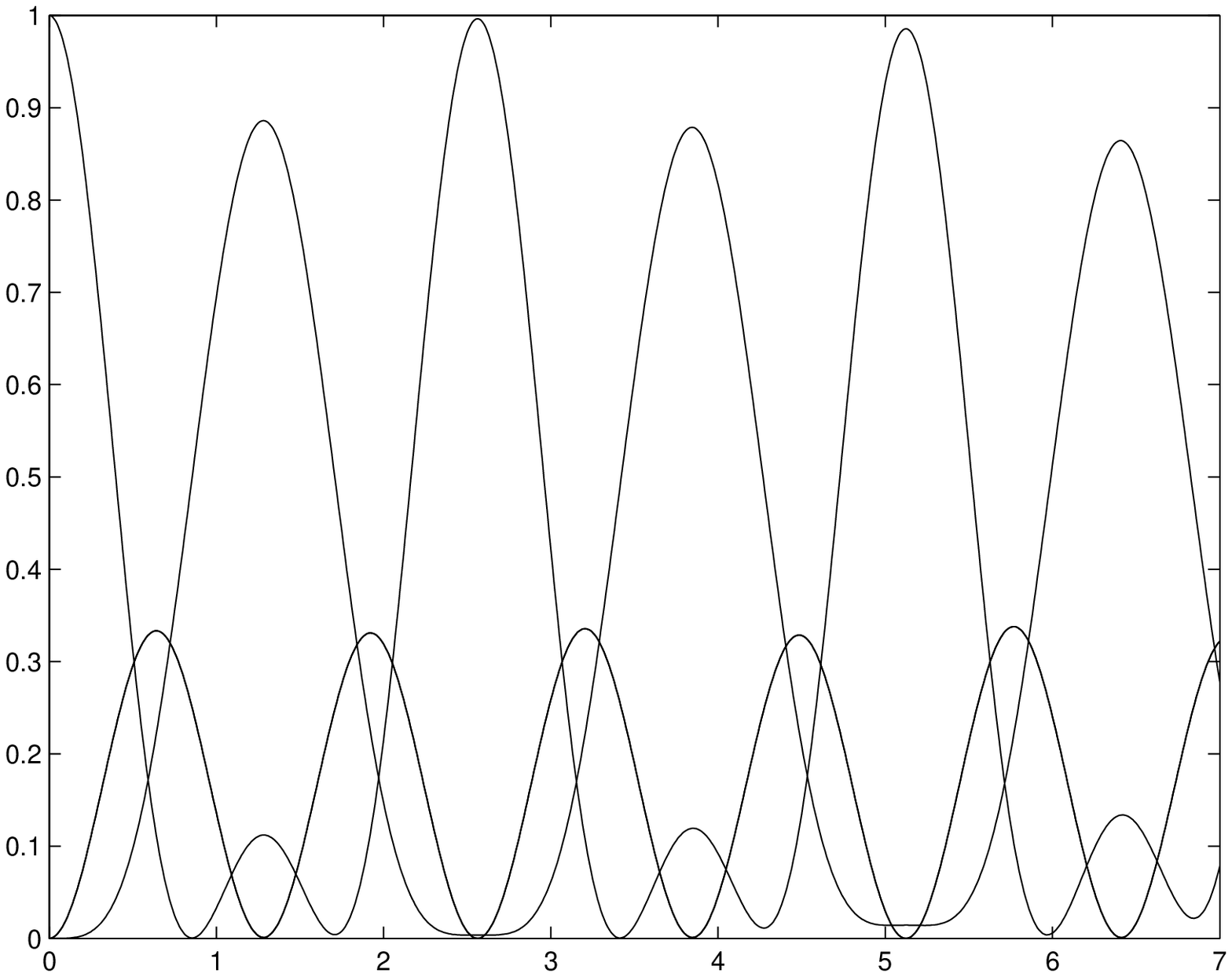} %
\includegraphics[width=18pc,height=12pc]{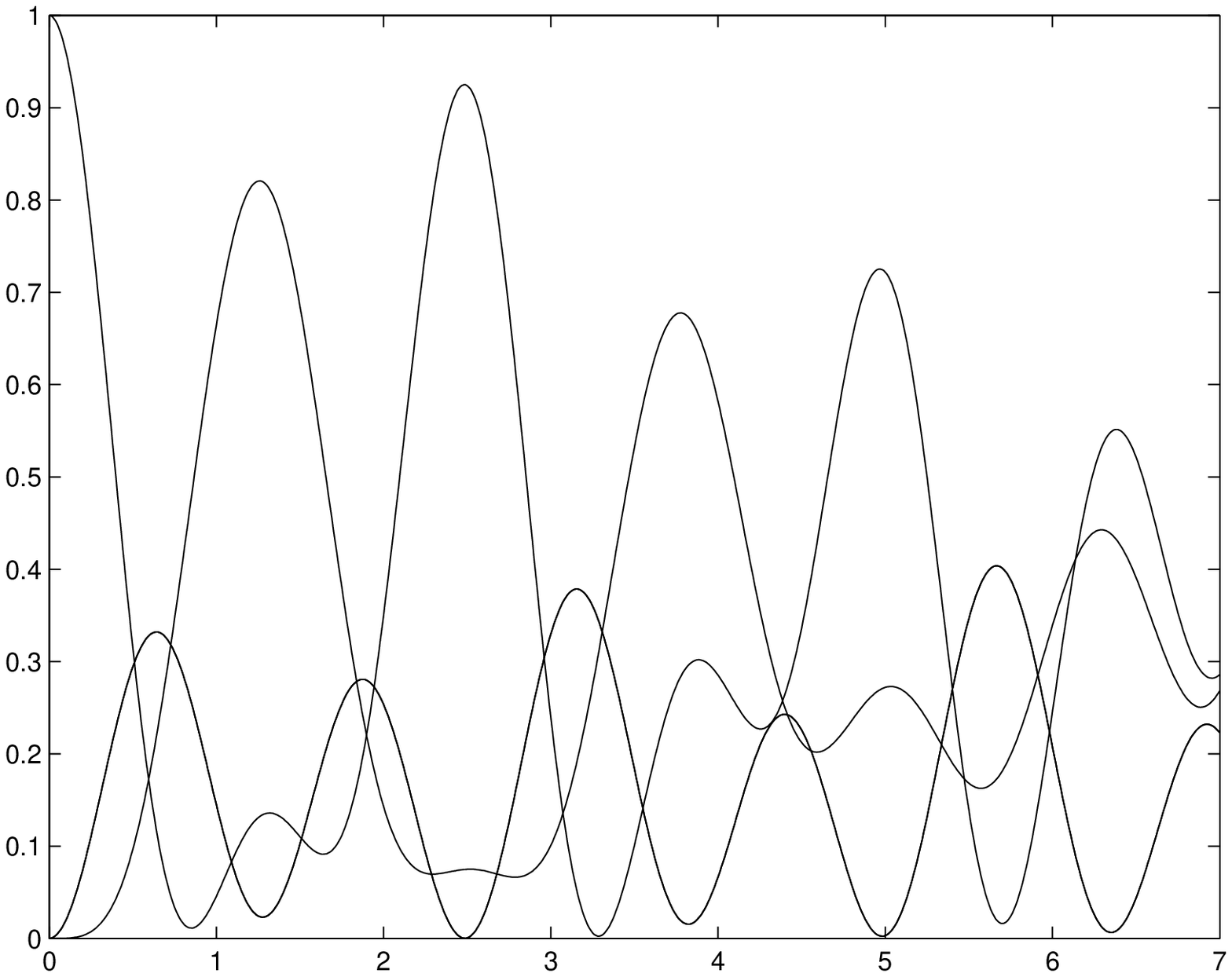} \put(-110,-10){$\tau$%
} \put(-325,-10){$\tau$}
\end{center}
\caption{The population for the atomic system initially in excited state
with $n=0$ and (a) $\Delta =0.1$ (b) $\Delta =0.5$.}
\end{figure}
As the time goes on another partially entangled state is generated around $%
\tau \simeq 0.5$. This state could be written as superposition of
the ground and excited state in additional to a separable part,
this classes is similar to Werner states \cite{Wer,Englert}. In a
mathematical form one writes this class as
\begin{equation}
\bigl|\psi _{3}\bigr\rangle=\eta (\bigl|ee\bigr\rangle+\bigl|gg\bigr\rangle%
)+\zeta \bigl|eg\bigr\rangle
\end{equation}%
This state is periodically appears round $\tau \simeq 2,3.5,...$.
Also, this class of states is similar to what is called Werner
state, which consists of fraction of maximum entangled state in
additional to completely mixed state On the other hand, there are
some cases where the entangled system turns into a product system.

In Fig.(1b), we set $\Delta=0.5$, in this case there are different classes
of entangled state are generated in additional to the classes which has been
appeared in Fig.(1a). As an example round $\tau\simeq 4.5$, there is a class
of entangled state can be written as
\begin{equation}
\bigl| \psi_4 \bigr\rangle\simeq \mu_2(\bigl| ee \bigr\rangle+\bigl| gg %
\bigr\rangle)+\nu(\bigl| eg \bigr\rangle+\bigl| ge \bigr\rangle)
\end{equation}
It is clear as one increases the detuning the atomic system remains
entangled for any $\tau>0$.

To see the effect of the initial state setting, we consider the case where
the atomic system is prepared initially in ground state i.e $\bigl| %
\psi(0)_a \bigr\rangle=\bigl| gg \bigr\rangle$. The behavior of this
phenomena is seen in Fig.(2). In this figure the most interesting remark,
that the generated entangled state between the two qubits is almost
entangled at any time $\tau>0$. Also, there are some different classes of
entangled states appear at different $\tau$. As an example, round $%
\tau\simeq 1.5$ the generated entangled state takes the form,
\begin{equation}
\bigl| \psi_5 \bigr\rangle=\chi_1\bigl| ee \bigr\rangle+\chi_2\bigl| gg %
\bigr\rangle+\chi_3(\bigl| eg \bigr\rangle+\bigl| ge \bigr\rangle).
\end{equation}
\begin{figure}[tbp]
\begin{center}
\includegraphics[width=18pc,height=12pc]{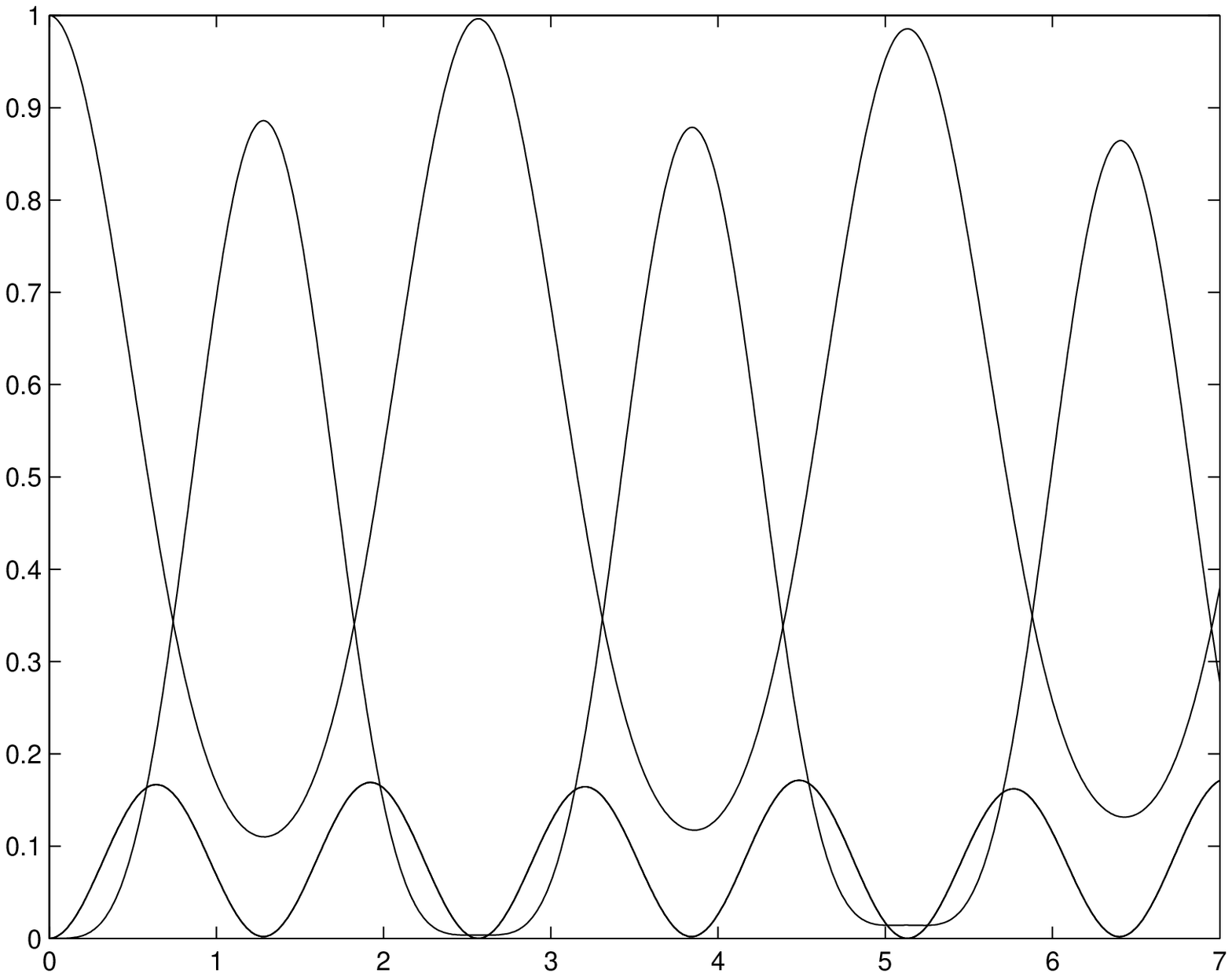} %
\includegraphics[width=18pc,height=12pc]{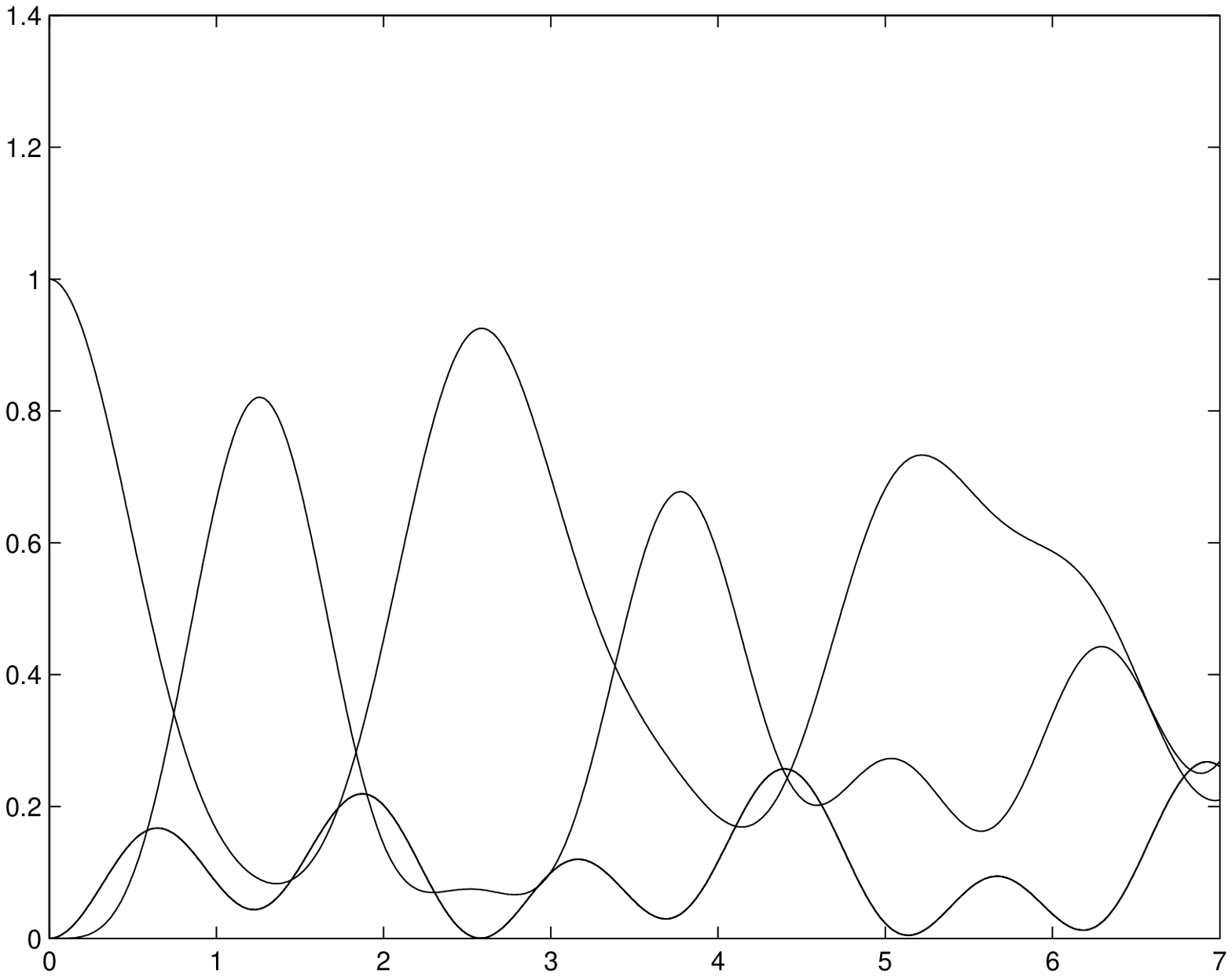} \put(-110,-10){$\tau$%
} \put(-325,-10){$\tau$}
\end{center}
\caption{The same as Fig.(1) but the atomic system is prepared initially in
ground state, $\bigl| \protect\psi(0)_a \bigr\rangle=\bigl| gg \bigr\rangle$%
. }
\end{figure}
This behavior is clear from Fig.(2a). For large values of the detuning the
same behavior is seen as shown in Fig.(1b). Also, the same classes of the
entangled states which has been depicted in Fig.(1a) appear clearly in
Fig.(2b) at different values of $\tau$ and weight.

\begin{figure}[tbp]
\begin{center}
\includegraphics[width=18pc,height=12pc]{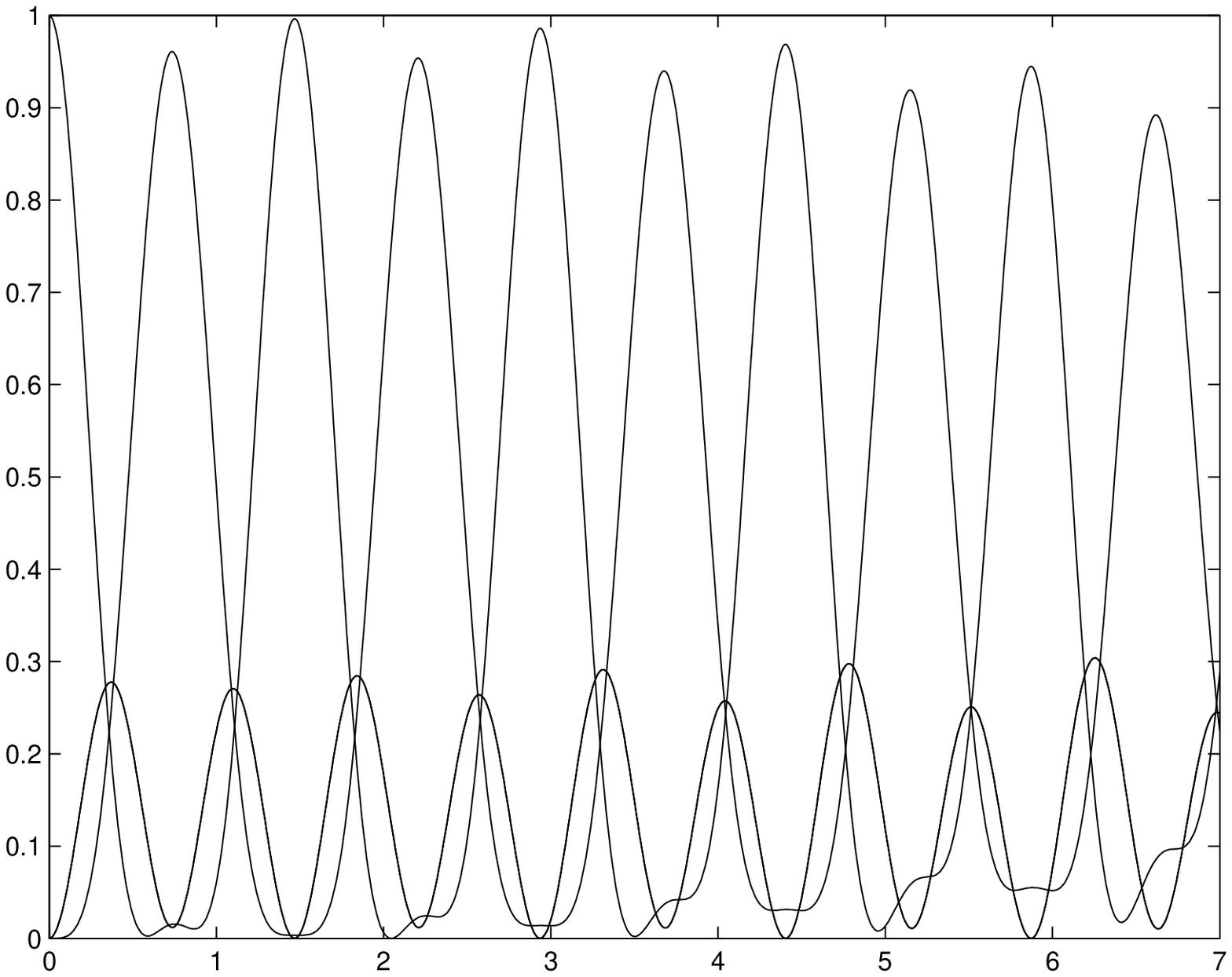} %
\includegraphics[width=18pc,height=12pc]{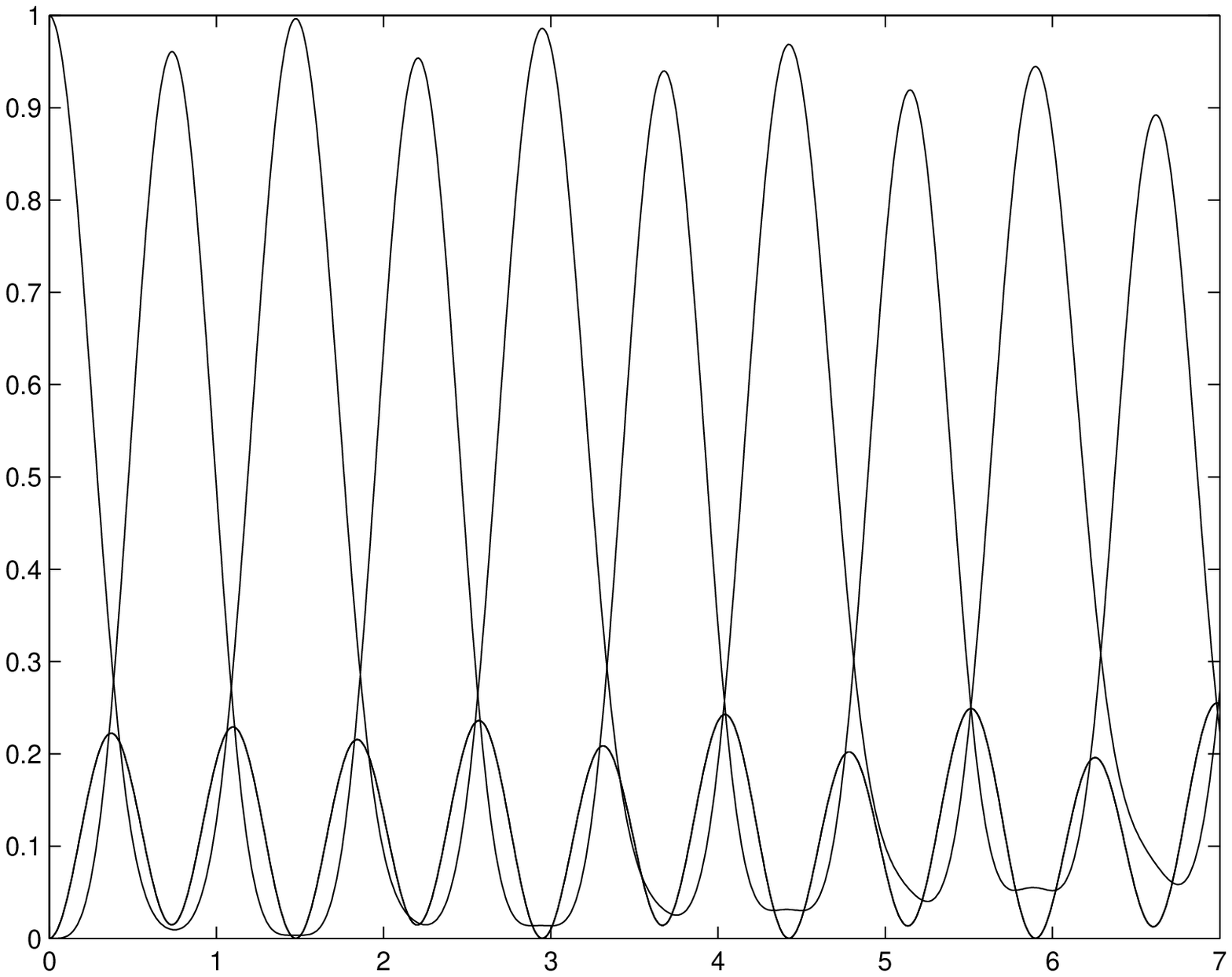} \put(-110,-10){$\tau$%
} \put(-325,-10){$\tau$}
\end{center}
\caption{The population for $\Delta =0.5$ and $n=3$. (a) The system is
initially prepared in excited state $\bigl|\protect\psi (0)_{a}\bigr\rangle=
\bigl|ee\bigr\rangle$ (b) The atomic system is prepared in ground state $
\bigl|\protect\psi (0)_{a}\bigr\rangle=\bigl|gg\bigr\rangle$.}
\end{figure}
Finally, we investigate the effect of the number of photon inside the cavity
on the populations dynamics. Fig.(3), shows this behavior for different
initial states setting, excited or ground state. The general behavior as
expected is that the number of oscillations is increased. This is clearly
seen by comparing Fig.(1b) and Fig(3), for a system prepared in excited
state and Fig.(2b) and (3b), for atomic system prepared in the ground state.
Also, the same classes of the generated entangled states are seen in this
case at different values of $\tau $. But these entangled state has a large
weight, so the degree of entanglement as we shall see in the next section is
much greater.

So, one can conclude that, to generate different classes of entangled states
between two qubits, one has to increase the number of photon inside the
cavity. Also, by increasing the detuning the generated states are almost
entangled for any time $\tau >0$.

\section{Entanglement}

We have seen in the previous section, there are some entangled state
generated in some interval of time. In this section, we quantify the amount
of entanglement contained in theses states. To quantify the amount of
entanglement contained in the entangled states, we shall use a measurement
introduced by K. Zyczkowski \cite{Zyc}. This measure states that if the
eigenvalues of the partial transpose are given by $\mu_i, i=1,2,3,4$, then
the degree of entanglement is defined by
\begin{equation}
\mathcal{N}=\sum_{i=1}^{4}{|\mu_i|-1}
\end{equation}
It is clear that the $\mathcal{N}$ is equal to zero for separable states and
it is equal to one for the maximal entangled states. Figure $2$, shows the
behavior of the $\mathcal{N}$ as a function of the scaled time $\tau$.

Fig.(4) shows the dynamics of the degree of entanglement for atomic system
is prepared initially in excited state. Fig.(4a), is plotted for $n=0$ and $%
\frac{\Delta}{\lambda}=0.1$. It is clear as soon as the interaction goes on
an entangled state is generated. At $\tau=0$, the degree of entanglement $%
\mathcal{N}=0$ and then increase until it reaches its maximum values then
reduces to zero, where the atomic system converted into a product state. As
time increases, this behavior is repeated, but the average amount of
entanglement is increased. In Fig.(4a),we consider the detuning parameter $%
\frac{\Delta}{\lambda}=0.5$, we can see that in this case, the generated
entangled state is much better and does not converted into a separable state
as the time increases. So, one can say that by increasing the value of the
detuning parameter, one can get an entangled state with long-lived
entanglement.

To see the effect of the initial state setting, we consider the atomic
system is prepared initially in the ground state. The dynamics of
entanglement for this case is depicted in Fig.(5). It is clear the same
behavior is seen as that depicted in Fig.(4), but the generated entangled
state is survival for a long time. This is clear if one compares Fig.(4a)
with Fig.(5a), where the degree entanglement vanishes round $\tau =0.8$, for
the first time while it vanishes round $\tau =1.6$ for a system is initially
prepared in the ground state. Also, as one increases the detuning parameter
the possibility of long-lived entanglement is much better. So, one of the
most better choice to generate en entangled state with high degree of
entanglement, is that the atomic system is prepared in ground state and a
large values of the detuning parameter \cite{aty07}.

Finally, we consider the effect of the number of photon inside the cavity $n
$, on the degree of entanglement. This behavior is depicted in Fig.(6),
where we consider $\frac{\Delta}{\lambda}=0.5$ and $n=3$. In Fig. (6a), we
assume that the atomic system is prepared in excited state $\bigl| ee %
\bigr\rangle\bigl\langle ee \bigr|$. From this figure, although the
generated entangled state is long-lived entanglement, but the degree of
entanglement $\mathcal{N}$, is smaller than that has been shown in Fig,(4a),
where the number of photon inside the cavity $n=0$. Also, as a remark, we
can see that the maximum entangled state is shifted to the right as the time
increases.

\begin{figure}[tbp]
\begin{center}
\includegraphics[width=18pc,height=12pc]{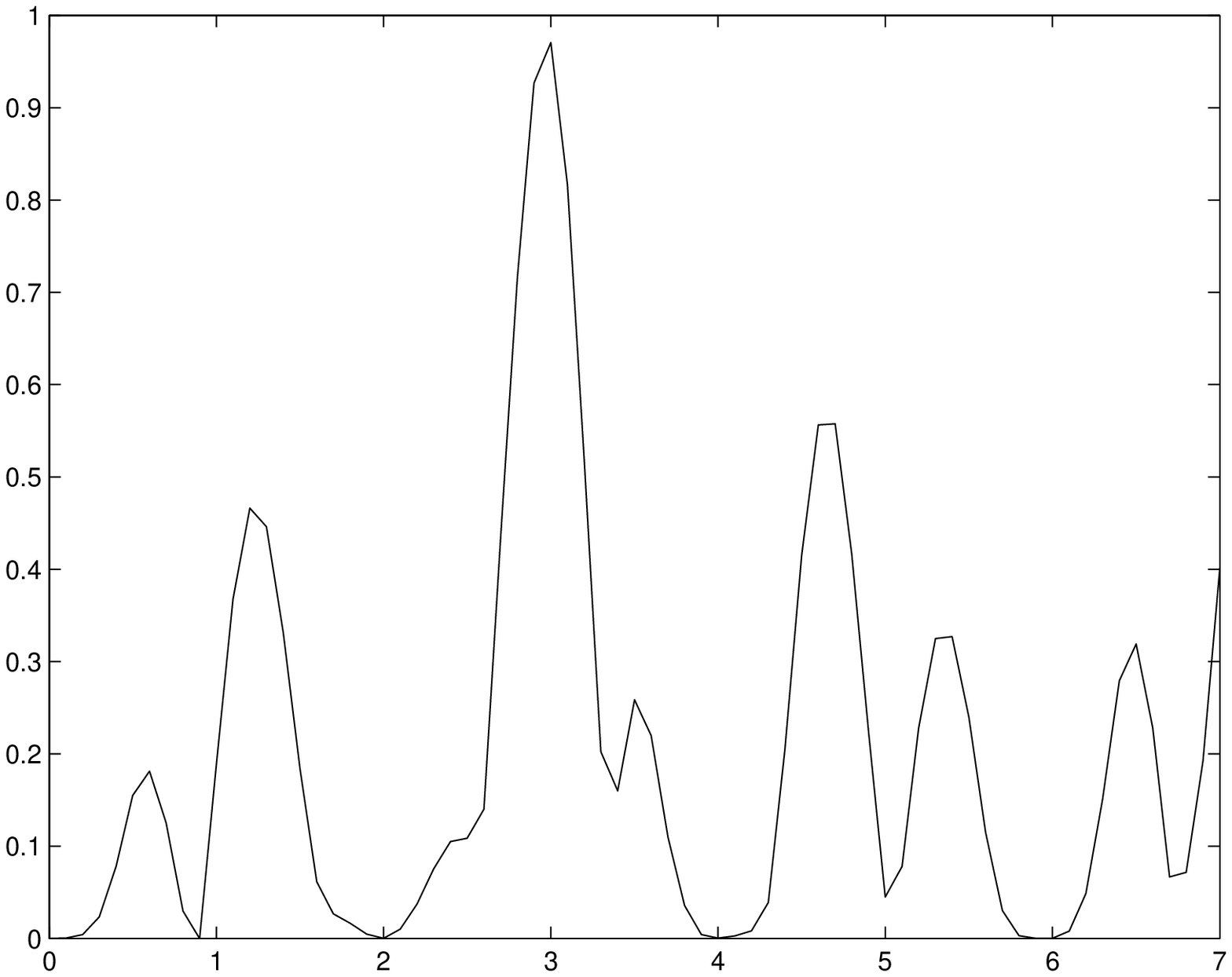}\ %
\includegraphics[width=18pc,height=12pc]{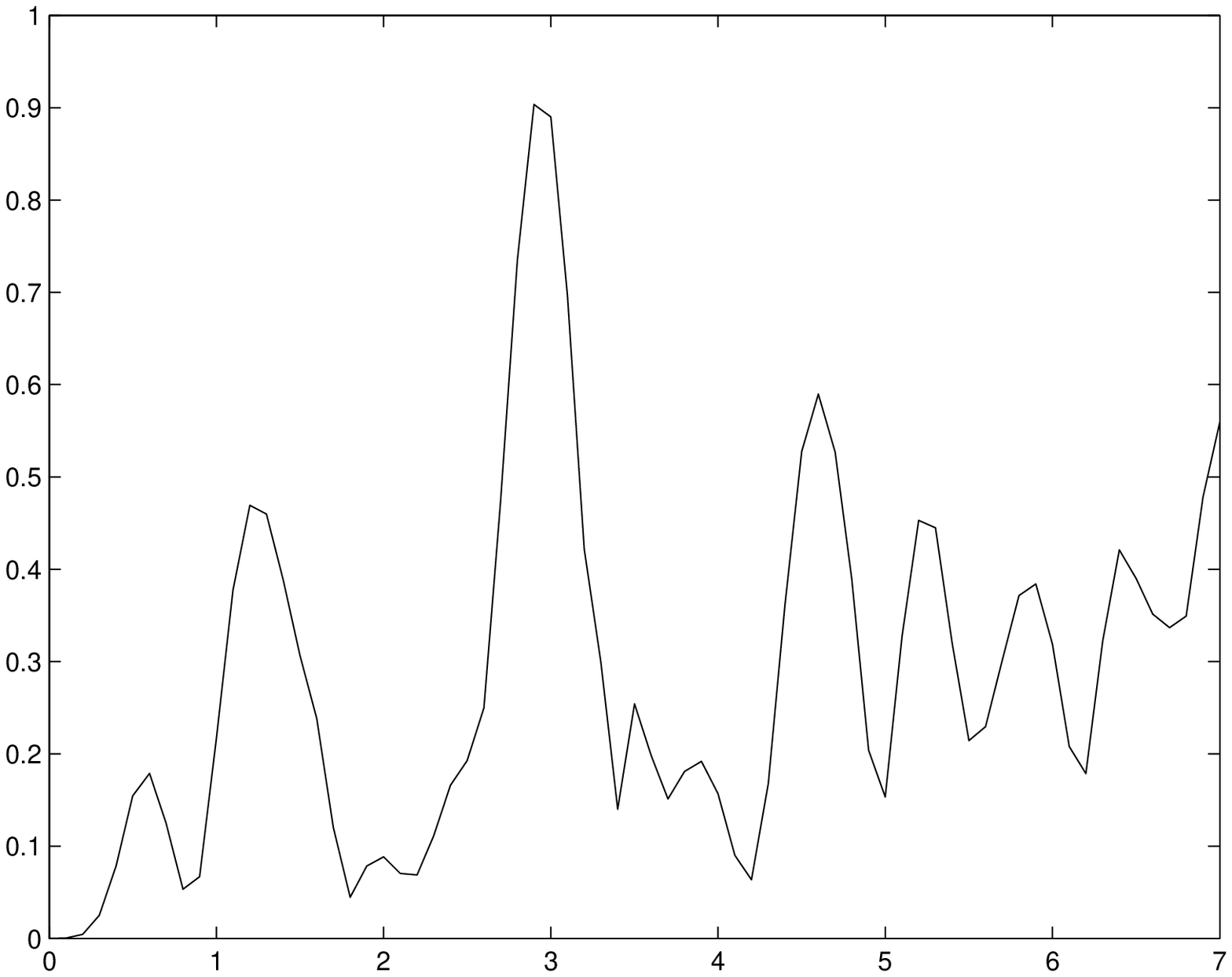} \put(-110,-10){%
$\tau$} \put(-325,-10){$\tau$}
\end{center}
\caption{The degree of entanglement for the atomic system is prepared
initially in excited state with $n=0$ and (a) $\Delta=1$ (b) $\Delta=0.5$.}
\end{figure}
\begin{figure}[tbp]
\begin{center}
\includegraphics[width=18pc,height=12pc]{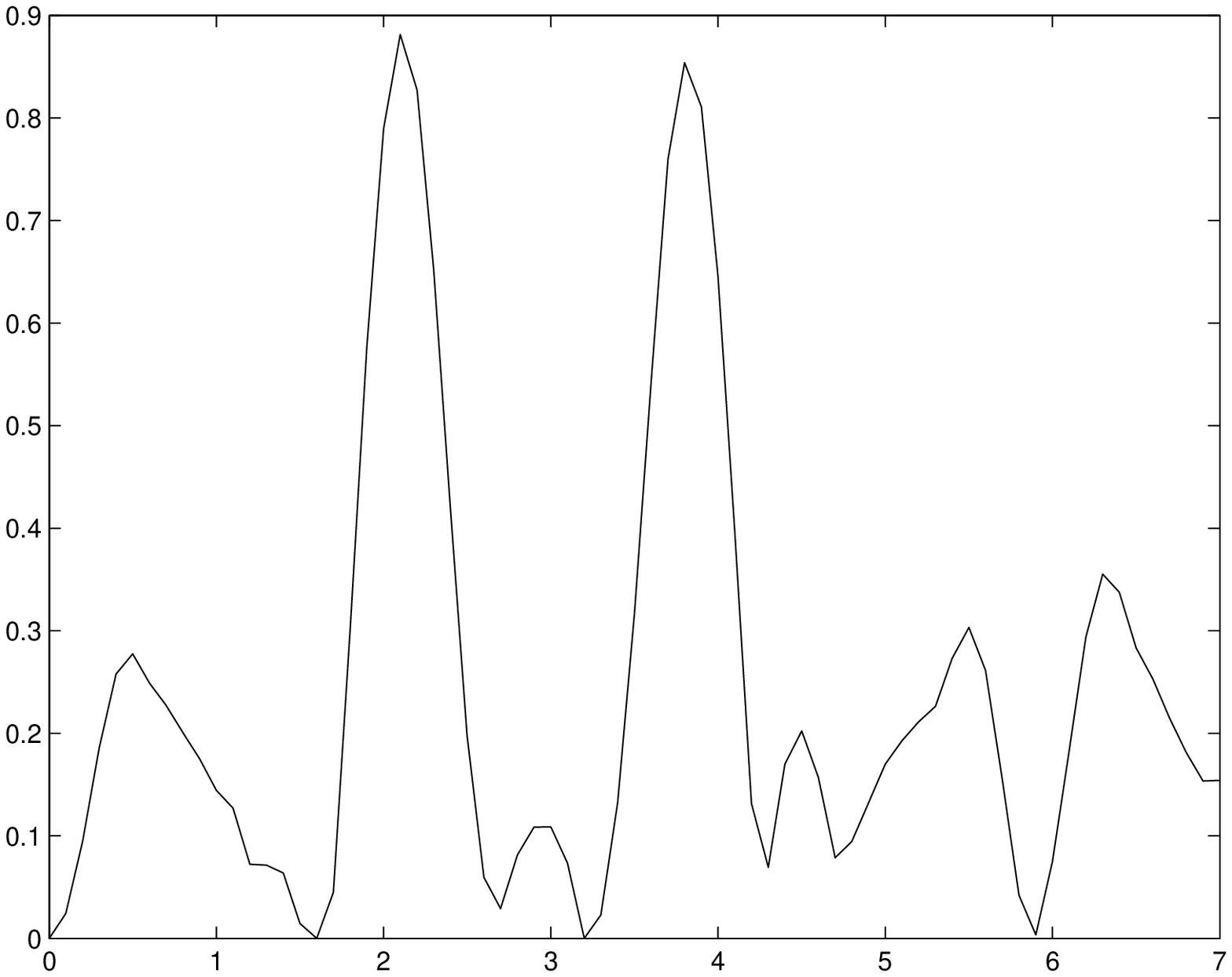}\ %
\includegraphics[width=18pc,height=12pc]{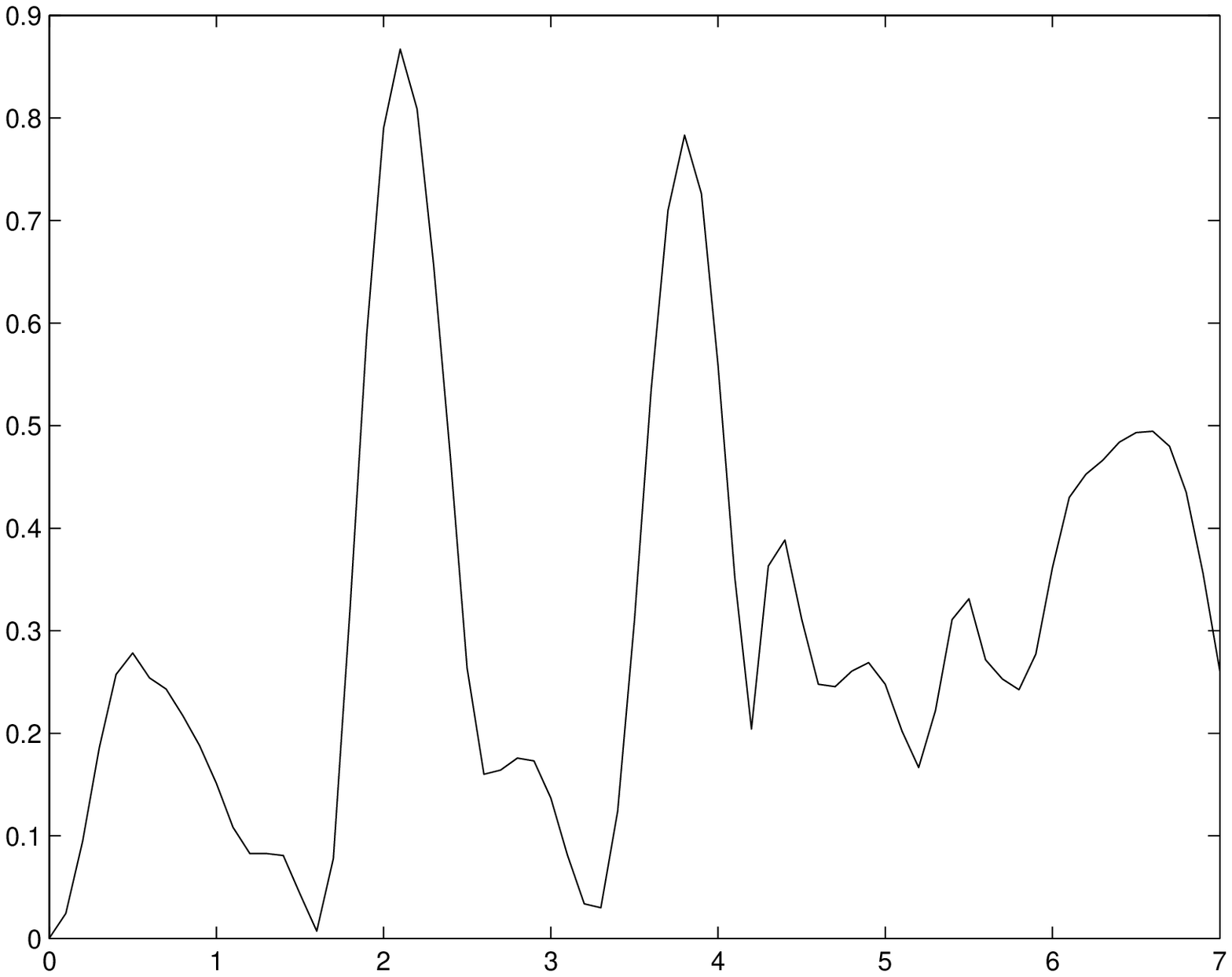} \put(-110,-10){%
$\tau$} \put(-325,-10){$\tau$}
\end{center}
\caption{The same as Fig.(1), but the system is prepared initially in the
ground state.}
\end{figure}

\begin{figure}[tbp]
\begin{center}
\includegraphics[width=18pc,height=12pc]{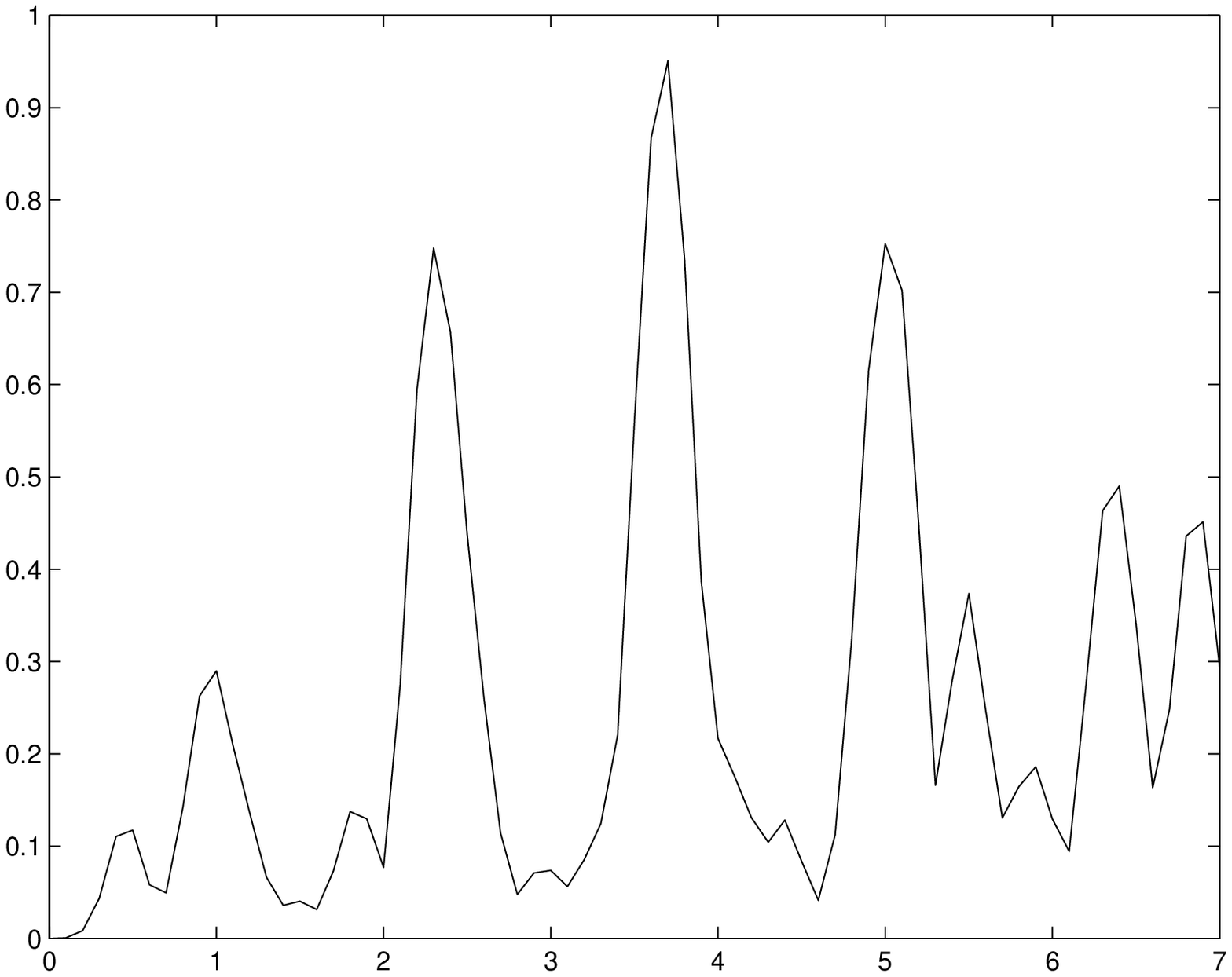} %
\includegraphics[width=18pc,height=12pc]{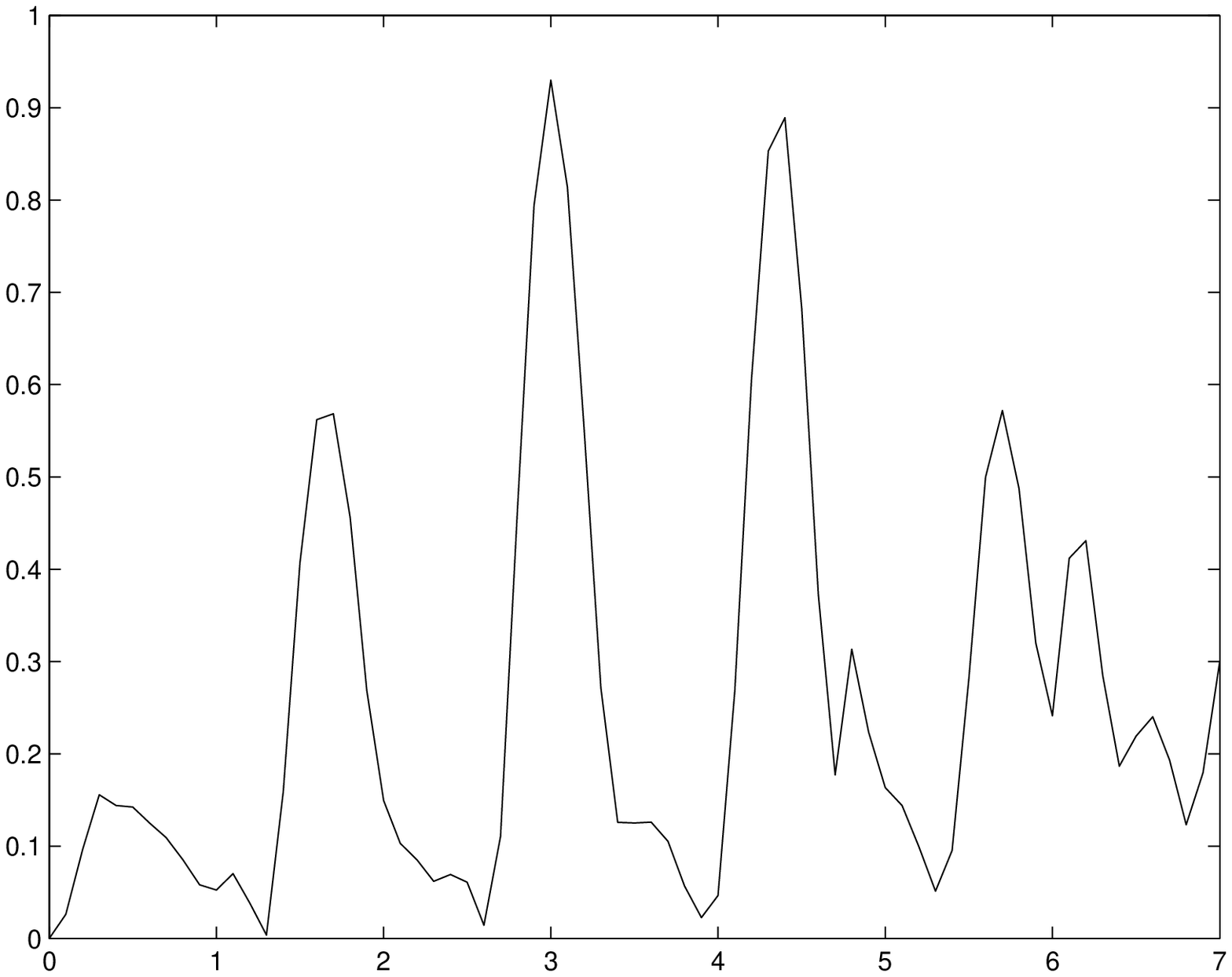} \put(-110,-10){%
$\tau$} \put(-325,-10){$\tau$}
\end{center}
\caption{The degree of entanglement where $\Delta=0.5, n=3$.(a) the system
is initially prepared in the excited state(b) The system is prepared in
ground state}
\end{figure}

\section{Conclusion}

In this contribution, we have investigated the dynamics of an atomic system
interacting with a cavity field. We considered different classes of the
atomic initial states and Fock state as an initial state of the field. The
effect of the number of photon as well as the detuning parameter on
generating entangled states is investigated in different regimes. We find
that there are several classes of entangled states can be generated. Some of
these classes are of Bell's state types, others are of Werner state types.
Also, it is shown that, the generation of entangled states with long lived
entanglement depends on the initial state setting. If the atomic system is
prepared in the ground state, the generated entangled state survived longer
than the case in which the system is prepared in an excited state. On the
other hand as one increase the numbers of photon inside the cavity, the
entangled state live longer although the amount of entanglement is smaller
in average. Finally, it is shown that the detuning parameter plays a central
role on controlling the entanglement, and the survivability of entanglement
increases once the detuning is increased. 

\end{document}